# A Model of Presidential Debates

Doron Klunover[a], John Morgan[b]

February 2020

## Abstract

Presidential debates are viewed as providing an important public good by revealing information on candidates to voters. We consider an endogenous model of presidential debates in which an incumbent and a challenger (who is privately informed about her own quality) publicly announce whether they are willing to participate in a public debate, taking into account that a voter's choice of candidate depends on her beliefs regarding the candidates' qualities and on the state of nature. It is found that in equilibrium a debate occurs or does not occur independently of the challenger's quality and therefore the candidates' announcements are uninformative. This is because opting-out is perceived to be worse than losing a debate and therefore the challenger never refuses to participate.

[a] Corresponding author: Department of Economics, Ariel University, 40700 Ariel, Israel. E-mail address: doronkl@ariel.ac.il.
[b] Haas School of Business and Department of Economics, University of California, Berkeley, California 94720.

1. Introduction

In US presidential elections, debates between the candidates are major media events. Even the least watched debate had an audience share of about 30 percent (Erikson and Wlezien, 2012).[1] The debates may or may not have a significant effect on voters (for evidence supporting the former view, see Geer, 1988; Bidwell et al., 2019; Brierley et al., 2020; for evidence supporting the latter view, see Abramowitz, 1978; Miller and MacKuen, 1979; Lanoue, 1991). Erikson and Wlezien (2012) point out that although there is some anecdotal evidence regarding presidential debates, their effect on voter behavior is hard to measure and therefore remains an open question. Furthermore, while in some countries, such as the US, debates are regularly held, in others, such as Israel, they are rare.[2] In particular, candidates are usually not obligated to participate in a debate and they take place only if both candidates agree to do so. In light of the aforementioned measurement difficulties, it would appear that a theoretical model is called for. In what follows, we build a game theoretic model that is, to the best of our knowledge, the first to evaluate the interactions between candidates and voters in presidential debates (although it may also apply in other contexts).

---

[1] The debates held in 2000 and 2004 were the least watched among all debates held prior to 2012.

[2] In Israel, one-on-one debates between the two leading candidates for prime minister were held regularly only between 1977 and 1996. Even in the US, presidential debates were not held between 1964 and 1972.

We essentially attempt to answer two important questions: 1) Under what conditions are debates held? and 2) Are they informative or noisy?

Specifically, we consider a model in which an incumbent and a challenger (who is privately informed about her own quality) are running for president. Before the elections are held, each candidate publicly announces whether she is willing to participate in a debate, in which the winner—from the voters' perspective—is stochastically determined according to the candidates' qualities. On Election Day, the (median) voter's choice of whom to vote for depends on the candidate's expected qualities and on nature, which is a random variable realized on Election Day.

We show that the game's equilibrium is independent of the challenger's quality. In particular, there exists a unique equilibrium in pure strategies in which the challenger always announces that she is willing to participate in a debate, whereas the incumbent's announcement depends on other fundamentals in the model. This can be viewed as "the dictatorship of the incumbent." Specifically, under reasonable conditions, the incumbent chooses to participate in a debate when her quality is low and not to participate when her quality is high. On the other hand, the challenger chooses to participate in a debate even when defeat is certain and, more generally, a challenger who participates and loses is always perceived to be a stronger candidate than one who opt-outs. This is because although beliefs about a candidate's quality are downgraded after a loss, the lower bound on these beliefs is still higher than the upper bound in the case that the challenger had opted-out. The results correspond to what is commonly observed in elections, namely that strong incumbents often choose to avoid debates.[3]

Given this behavior, the candidates' announcements are uninformative since no new information about the challenger's quality is revealed. However, if a debate is held, and depending on the shape of the probability distribution of the challenger's quality, the debate can be either informative or noisy. Furthermore, under reasonable conditions, the announcements made by the candidates are shown to be independent of the sequence in which they are made, and every sequence satisfies Perfect Bayesian Equilibrium when it is endogenously determined.

The paper proceeds as follows: In the remainder of this section, we review the related literature. Section 2 describes the model and section 3 presents the results. In section 4 the results are generalized to a game in which the sequence of announcements is endogenously determined. Section 5 concludes.

**Review of the literature**

The current model brings together two strands of the literature: information transmission and the effect of the media on voters in the context of a political campaign. More specifically, in the static version of our model, the problem of the incumbent can be viewed as a Bayesian persuasion problem (Kamenica

---

[3] Examples are presented following Proposition 1.

and Gentzkow, 2011). The incumbent who can be viewed as the sender shares the same prior with the voter who can be viewed as the receiver. She chooses the action that transmits a signal to the voter that maximizes her probability of being reelected.[4] Our result that the challenger's choice of whether or not to participate is not informative is related to the seminal paper of Grossman (1981) who shows that when a monopoly cannot reveal the quality of its product but can offer warranties against bad outcomes, then it will always offer a full warranty, regardless of the quality of the product. Therefore, in equilibrium the quality of the product is unrevealed. Note that apart from the context being different, in our model the choice of whether to participate in a debate transmits two different and sequential signals (the choice whether to participate followed by the outcome of the debate, if it occurs). As noted, the

---

[4] A related study of competition between multiple senders in the context of Bayesian persuasion is Gentzkow and Kamenica (2017). Another is Gul and Pesendorfer (2012) who consider two rival parties that provide costly information to a voter who then chooses between them.

key intuition behind our result is directly attributable to this sequential structure.[5]

A presidential debate is a major media event and therefore our paper is also related to the large empirical literature on the effect of the media on voters. In this literature, researchers have examined various aspects of effect and have arrived at differing conclusions. For instance, DellaVigna and Kaplan (2007) find that Fox news has a significant effect on voters. Gerber et al. (2011) find that televised campaign ads have a strong but short-lived effect on voter's preferences, while Spenkuch and Toniatti (2015) find that ads have no impact on turnout but do have a significant effect on voting. In the context of political debates, evidence appears to be mixed (as mentioned above). However, more recent studies show that debates are important. In particular,

---

[5] Note that a presidential debate is modeled here as a contest (see Dixit, 1987; Konrad, 2009). However, unlike the usual contest models, we assume that the outcome of a debate depends only on players' qualities rather than their effort. Although participating in a debate certainly requires a certain amount of effort, we nevertheless posit that the effort invested and its effect on the debate's outcome are of marginal importance and moreover the outcome is primarily determined by the candidates' inherent abilities. For studies that analyze how information about contestants is inferred from the contest outcome, see Krähmer (2007) and Noe (2020). For a recent study on signaling in contests, see Denter et al. (2018).

Bidwell et al. (2019) conducted a large-scale field experiment in Sierra Leone, where information on candidates is scarce. They concluded that political debates have a significant effect on voter opinion and therefore can also trigger a chain of events, such as an increase in campaign spending. Moreover, using a unique data set, Brierley et al. (2020) shows that debates have a strong effect on partisan voters.

Our main contribution to the existing literature is the analysis of presidential debates from a rational point of view, and although the analysis does not consider specific aspects of a debate, such as the appearance and verbal skills of the candidates,[6] it does help to understand the nature of a debate by demonstrating that the choice to participate is more important than the debate itself.

**2. The Model**

Two candidates, an incumbent and a challenger, are running for president. The incumbent's quality is commonly known to be $q_I(\in (0,1])$ while the challenger's quality, $q$, is a random variable with a probability distribution $p$ over the interval $[0,\infty)$ with mean $\bar{q} > 0$. The actual value of $q$, $q_C(\in [0,\infty))$ is private information known only to the challenger until it is learned by all after Election Day. Since the incumbent has already served as president while the challenger has not, we assume that there is more information available on the incumbent (which may not necessarily be the case). Furthermore, we restrict $q_I$ to be bounded from above in order to assign a positive probability to the case in which the challenger's quality is higher than that of the incumbent.[7]

---

[6] Lawson et al. (2010) examines this effect.

Before Election Day, each candidate publicly and simultaneously announces whether she is willing to participate in a debate (P) or not (NP), where the probability of the challenger winning the debate is $\theta(q_C, q_I)$ and that of the incumbent is $1 - \theta(q_C, q_I))$, where $\theta$ satisfies the usual properties:[8]

(1) $\frac{\partial \theta}{\partial q_I} < 0$, $\frac{\partial \theta}{\partial q_C} > 0$, $\frac{\partial^2 \theta}{\partial q_C^2} < 0$, $\theta(0, q_I) = 0$ and when $q_C \to \infty$, $\theta(q_C, q_I) \to 1$ $\forall q_I \in (0, 1]$.

A debate is held only when both candidates announce P.

---

[7] In principle, $q_I$ can be bounded from above by any real positive number and the specific assumption of a boundary of 1 is made without loss of generality.

[8] Note that the constraints on $\theta$ with respect to $q_C$ are tighter than those on $\theta$ with respect to $q_I$. Therefore, we do not require anonymity.

*2.1 The voter*

There is one voter. In the case that a debate is not held, she observes the candidate's announcements before the elections. In the case that a debate is held, she observes who won the debate.[9]

If we define $\bar{q}^{el}$ to be the expected value of $q$ on Election Day, then the voter chooses the challenger when $\bar{q}^{el} - q_I > \varepsilon$, where $\varepsilon (\in R)$, which represents nature, is a random variable independent of $q$ with a commonly known cumulative distribution G that is realized on Election Day (before the voter chooses a candidate). Note that $\varepsilon$ can be viewed as a measure of the match—from the voters' perspective—between the incumbent and the state of nature.[10] For instance, if a natural disaster takes place in the time between the debate and the elections and the voter has determined the expected quality of the challenger to be equal to that of the incumbent, then the voter may vote for the more experienced candidate (corresponding to the case in which the realization of $\varepsilon$ is positive). More generally, a candidate usually has characteristics other than quality that may give her an advantage in certain scenarios. Note that the voter only cares about the expectation of the challenger's quality (not about its distribution) and therefore she is risk-neutral.

---

[9] Note that officially there are no winners and losers in presidential debates. However, after a debate is held, there is usually a consensus as to who the "winner" is. For instance, after the Democratic debate on January 14, 2020, Matthew Yglesiasis wrote the following in a Vox column: "Joe Biden walked onto another Democratic debate stage Tuesday night as the frontrunner and once against walked off the winner by default."

[10] He can be viewed as the "median voter". In that case, in any realization of $\varepsilon$ in which $\bar{q}^{el} - q_I < \varepsilon$, the majority of voters prefer the incumbent and therefore vote for her.

After the announcements are made and the debate is held (or not), and before $\varepsilon$ is realized, the probability of the challenger winning the election is therefore $G(\bar{q}^{el} - q_I)$.[11]

*2.2 Timeline*

In view of the above, the timeline of the model is as follows:

1. Candidates simultaneously choose P or NP.
2. A debate is held iff both candidates choose P.
3. $\varepsilon$ is realized.
4. The voter chooses a candidate.

After the announcements are made, the voter follows a decision rule and hence is not defined here as a player. We therefore consider a Bayesian game with two players (i.e. the incumbent and the challenger) who have the same choice set: $\{P, NP\}$, but different information sets. We therefore focus our attention on Bayesian equilibria.[12]

---

[11] Dixit (1987) examines a similar framework for noisy contests.

*2.3 Candidates*

Each candidate maximizes her expected probability of winning the election. We assume that a candidate is indifferent between winning with a certain probability and winning with an expected probability equal in value. Therefore, both the voter and the candidates are risk-neutral. In the remainder of this section, we describe each candidate's problem and add some notation that will be useful in the remainder of the analysis.

***The challenger.*** Let $E[G|q_C]$ be the challenger's expected probability of winning the election from her perspective (she knows her own quality) after the announcements are made and before the debate is held (or before Election Day if the debate is not held).

Note that if a debate is held, $\bar{q}^{el}$ is determined following it. Otherwise, $\bar{q}^e$ is determined right after the announcements are made, in which case $E[G|q_C] \equiv G(\bar{q}^{el} - q_I)$.

---

[12] In section 4, we consider a dynamic extension of the model, in which the sequence of announcements is endogenously determined. There we consider Perfect Bayesian Equilibria.

In particular, if only the challenger announces NP, let $\bar{q}^{el} \equiv \bar{q}|C_{NP}$ and if only the incumbent announces NP, let $\bar{q}^{el} \equiv \bar{q}|I_{NP}$; if both announce NP, let $\bar{q}^{el} \equiv \bar{q}|IC_{NP}$ and if both announce P, let $E[G|q_C] \equiv E[G|q_C]|D$.

Let $p(q')|w$ be the probability that the challenger's quality is $q'$ in the case that a debate was held and she won, and let it be $p(q')|l$ in the case that a debate was held and she lost, for all $q' \in [0,\infty)$. Also, let $\bar{q}^{el} \equiv \bar{q}|w$ if the challenger won and let $\bar{q}^{el} \equiv \bar{q}|l$ if she lost. It then follows that:

(2) $E[G|q_C]|D$

$= \theta(q_C, q_I)G\big((\bar{q}|w) - q_I\big) + (1 - \theta(q_C, q_I))G((\bar{q}|l) - q_I)$.

Given the incumbent's announcement, the challenger makes the announcement that maximizes $E[G|q_C]$. Formally, given the incumbent's announcement, the challenger's maximization problem is:

(3) $\max\limits_{\text{announcement} \in \{P, NP\}} E[G|q_C]$.

*The incumbent.* Let $EG$ be the challenger's expected probability of winning the election from the incumbent's perspective (since she does not observe the challenger's quality) after announcements are made and before the debate is held (or before Election Day in the case that the debate is not held).

In the case that both candidates announce P, let $EG \equiv EG|D$.[13] Otherwise, by definition, $EG \equiv G(\bar{q}^{el} - q_I) \equiv E[G|q_C]$.

---

[13] It is shown later that $EG|D$ is identical to (6).

Given the challenger's announcement, the incumbent makes the announcement that maximizes her expected probability of winning the election, i.e., $1 - EG$, which therefore minimizes $EG$. Formally, given the challenger's announcement, the incumbent's minimization problem is:

(4) $\min_{\text{announcement} \in \{P, NP\}} EG.$

We now proceed to the analysis of the model's results.

## 3. Results

*3.1 Mandatory participation in the debate*

It is useful at this point to consider the case in which participation in the debate is mandatory (i.e., neither of the players needs to make a decision).

We add the lower index $m$ to all notations in this subsection. For instance, $\bar{q}|w_m$ is the challenger's expected quality given that she wins the debate, and $\bar{q}|l_m$ is her expected quality given that she loses.[14] It follows that:

(5) $E[G|q_C]|D_m$

$= \theta(q_C, q_I)G((\bar{q}|w_m) - q_I) + (1 - \theta(q_C, q_I))G((\bar{q}|l_m) - q_I)$

and

(6) $EG|D_m$

---

[14] Note that $\bar{q}|w_m = \frac{\int_{q=0}^{\infty} p(q)\theta(q,q_I)q\,dq}{\int_{q=0}^{\infty} p(q)\theta(q,q_I)\,dq}$ and $\bar{q}|l_m = \frac{\int_{q=0}^{\infty} p(q)(1-\theta(q,q_I))q\,dq}{\int_{q=0}^{\infty} p(q)(1-\theta(q,q_I))\,dq}$. For detailed calculations, see equations (7)-(10) in the appendix.

$$=$$

$$(\int_{q=0}^{\infty} p(q)\theta(q,q_I)\,dq)G((\overline{q}|w_m) - q_I) +$$

$$(\int_{q=0}^{\infty} p(q)(1 - \theta(q,q_I))dq)G((\overline{q}|l_m) - q_I).$$

We now present a technical lemma that will be useful in the remainder of the analysis.

*Lemma 1*

(i) $\overline{q}|w_m > \overline{q} > \overline{q}|l_m$ and therefore, $E[G|0]|D_m < G(\overline{q} - q_I)$, and $E[G|q_C]|D_m > G(\overline{q} - q_I)$ for a sufficiently large $q_C$.

(ii) $EG|D_m < (>)G(\overline{q} - q_I)$ when $G$ is concave (convex) over the interval $[(\overline{q}|l_m) - q_I, (\overline{q}|w_m) - q_I]$.[15]

All proofs appear in the appendix. We now proceed to analyze the equilibrium of the original game.

*3.2 Equilibrium*

Notice that the voter and the incumbent share the same information set and therefore the incumbent's announcement cannot reveal information to the voter. Regarding the challenger's announcement, if the incumbent announces P, then the challenger's announcement is decisive (i.e., her announcement determines whether or not a debate will be held) and therefore it may reveal information to the voter, since it is made given a specific $q_C$ that is privately

---

[15] See the discussion of the interval $[(\overline{q}|l_m) - q_I, (\overline{q}|w_m) - q_I]$ following Proposition 1.

known to the challenger. Otherwise (i.e. in the case that the incumbent announces NP), the announcement made by the challenger is not decisive and therefore does not reveal information. We can therefore state the following:

**Fact 1** *Given that the incumbent announces NP, the probability distribution of q remains p and therefore* $\overline{q}|I_{NP} = \overline{q}|IC_{NP} = \overline{q}$.

Furthermore, we can state the following lemma:

*Lemma 2* *If the incumbent announces P, then so does the challenger.*

The key intuition behind Lemma 2 is that given the incumbent's announcement of P, an announcement of NP by the challenger implies that her quality is lower than if she had announced P and therefore the challenger always responds to P by also announcing P.

To see this, consider the following: Assuming that some challengers respond with P and others with NP and given that $G\big((\overline{q}|C_{NP}) - q_I\big)$ is independent of $q_C$ while $E[G|q_C]|D$ is continuous in $q_C$, there must exist a challenger who is indifferent between P and NP, and since $E[G|q_C]|D$ is monotonically increasing in $q_C$ all challengers who are stronger than this one will respond with P while weaker ones will respond with NP; however, in that case, the weaker ones will mimic the behavior of the stronger ones in order to be viewed as strong. This implies that the challenger's strategy can be consistent with beliefs only in a pooling equilibrium in which all challengers behave identically.

There are two additional and important points that should be clarified regarding the challenger's behavior. First, a belief that the challenger always responds to P with NP is not consistent with the challenger's strategy. To see this, assume that this is indeed the case. Then, the probability distribution of $q$ remains $p$ when the challenger chooses NP. However, since $E[G|q_C]|D$ is monotonically increasing in $q_C$, after a challenger deviates and responds to P with P, the updated beliefs about this challenger's quality must be such that either the probability distribution of $q$ remains $p$ or the left tail is truncated; either way, given these beliefs an extremely strong challenger prefers to respond to P with P, which is inconsistent with the initial belief that a challenger always responds to P with NP.

The second important point is that according to Lemma 2 the probability distribution of $q$ remains $p$ when the challenger responds to P with P. But what happens if she deviates and responds with NP instead? Then, by the same reasoning as above, the updated beliefs about this challenger's quality are such that the probability distribution of $q$ remains $p$ or the right tail is truncated. Either way, given these beliefs an extremely strong challenger will prefer to respond to P with P and therefore she is not the challenger who deviates by choosing NP, implying that there is a separating equilibrium and, as noted above, this is impossible.

Finally, note that Fact 1 together with Lemma 2 implies that, in all possible equilibria, the probability distribution of $q$ remains $p$ after announcements are made, and therefore the candidates' announcements are uninformative.

We are now in a position to characterize the game's equilibria.

***Proposition 1*** *The unique Bayesian equilibrium of the presidential debate game is as follows:*

(i) *A debate is held when $EG|D_m < G(\overline{q} - q_I)$.*

(ii) *A debate is not held when $EG|D_m > G(\overline{q} - q_I)$.*[16]

Proposition 1 implies that the equilibrium outcome is independent of the challenger's quality (i.e., $q_C$) and at the same time coincides with the incumbent's preferences. The challenger always announces P, and the choice of the incumbent, who makes the announcement that maximizes her expected winning probability in the election, is therefore always decisive.

Note that the effect of the incumbent's quality on the equilibrium outcome depends on the probability distribution of ε. In particular, in the case that ε has a unimodal probability distribution, $G$ is convex up to a certain point on the X-axis, after which it becomes concave. Therefore, since the interval $[(\overline{q}|l_m) - q_I, (\overline{q}|w_m) - q_I]$ "moves right" on the X-axis when $q_I$ decreases, by Lemma 1ii a decrease in the incumbent's quality can induce her to participate in the debate. For example, when ε has a normal distribution and $q_I$ goes to zero, this interval sits on the positive side of the X-axis where G is concave and therefore the incumbent announces P. It follows that a high-quality incumbent prefers to avoid a risky debate, while a low-quality incumbent may take her chances and participate.

---

[16] Specifically, the incumbent announces NP and the challenger announces P.

There is some anecdotal evidence to support this result. For example, the first presidential debate in US history between the two leading candidates was held in 1960, while there were no debates in the subsequent three elections between 1964 and 1972. In particular, in 1964 President Lyndon Johnson refused to participate in a debate and eventually won the election nonetheless. In 1968, Johnson did not run again and therefore the race was between two challengers, one of whom was Richard Nixon who won the election that year. In 1972, Nixon refused to participate in a debate and again won the election. Therefore, in the only two cases in US history in which the incumbent refused to participate he was later reelected.

In Israel, incumbent Prime Minister "Bibi" Netanyahu, who was clearly the favorite in the 2015 election and indeed won it, refused to participate in a debate against the challengers (eventually a debate was held between most of the challengers but without Netanyahu). In fact, Netanyahu participated in only one debate, which took place prior to the 1996 election in which he was first elected prime minister. He did not agree to participate in any debates subsequent to that one.[17]

Table 1 summarizes the expected probabilities of winning for the incumbent and the challenger, given the information available to them:

**Table 1**

| Incumbent/ Challenger | announce P | announce NP |
|---|---|---|
| announce P | $1 - EG\|D_m, E[G\|q_c]\|D_m$ | $1-G((\overline{q}\|C_{NP}) - q_I), G((\overline{q}\|C_{NP}) - q_I)$ |
| announce NP | $1 - G(\overline{q} - q_I), G(\overline{q} - q_I)$ | $1 - G(\overline{q} - q_I), G(\overline{q} - q_I)$ |

Each cell in the matrix contains the pair: $1 - EG$ and $E[G|q_c]$.[18]

*3.3 Information analysis*

In this section, we derive the sufficient conditions for a debate to be either informative or noisy.

Formally, we define a debate as informative (noisy) when $|E[G|q_C]|D - G(q_C - q_I)| < (>)|G(\overline{q} - q_I) - G(q_C - q_I)|$. Loosely speaking, in the case that a

---

[17]An exception is the 1999 election. Prior to it, Netanyahu participated in a debate against "Itzik" Mordechai. However, the main challenger, Ehud Barak, did not participate in that debate (but nonetheless won the election).

[18] Note that by Lemma 2 and the proof of Proposition 1, $G((\overline{q}|C_{NP}) - q_I) < E[G|q_c]|D_m$ and $\overline{q}|C_{NP} < \overline{q}$. The explicit form of $E[G|D_m]$ appears in the appendix in (13).

debate is held, if a candidate's expected probability of winning the election is the best available predictor of the corresponding probability in the case that $q_C$ is commonly known, then the debate is considered to be informative. Otherwise, it is noisy.[19]

In Lemma 3, we derive sufficient conditions for a debate to be either informative or noisy.

*Lemma 3*

    *(i)    A debate is informative if $q_C$ is sufficiently far away from $\bar{q}$.*

    *(ii)    A debate is noisy if $q_C$ is in the neighborhood of $\bar{q}$, and $\bar{q}$ is sufficiently small or large.*

Given that $q_C$ is unobservable, Lemma 3 implies that whether a debate is expected to be informative or noisy depends on the shape of the (commonly known) probability distribution of the challenger's quality. In particular, a debate is expected to be informative when voting for the challenger may look like a risky gamble, since it is most likely that she is either weak or strong. Alternatively, a debate is expected to be noisy when it is likely that the challenger is a weak candidate (or a strong one).

### 4. Endogenous sequence of announcements

---

[19] Note that this definition takes into account only the expected probability of winning and is not sensitive to its variance.

In this section, we allow for the sequence of announcements to be endogenously determined. In particular, the candidates first agree on the order of the announcements, and then each candidate makes her announcement in the agreed-upon order. This leads to the following proposition:

**Proposition 2** *If the sequence of announcements is endogenously determined, then there exist Perfect Bayesian equilibria that are characterized as follows:*

(i) *The probability distribution of q is believed to be p after the sequence of announcements has been agreed upon, regardless of which sequence that is.*

(ii) *Any sequence of announcements satisfies Perfect Bayesian Equilibrium.*

(iii) *The candidates' announcements are the same as in Proposition 1.*[20]

Note that in the equilibria described in Proposition 2, the candidates' expected probability of winning the election, the information available on Election Day and whether or not a debate will be held are all independent of the sequence of the announcements. The intuition behind this result is as follows: Let the sequence of announcements be exogenously determined. Since the preferences of the incumbent are commonly known, the challenger's announcement is decisive and therefore she cannot avoid a debate without being considered to be a weak candidate even when moving first. Furthermore, given that the challenger's expected probability of winning is the same for all sequences, if it is believed that the probability distribution of $q$ remains $p$ after the sequence has been agreed upon, then those beliefs are consistent with a challenger's strategy of approving any sequence of announcements that is proposed.

---

[20] Note that there are multiple pairs of announcements that satisfy a Perfect Bayesian Equilibrium when a debate is not held. In particular, any pair of announcements except (P,P) satisfies a Perfect Bayesian Equilibrium when the challenger makes the first announcement, and both (NP,NP) and (NP,P) satisfy a Perfect Bayesian equilibrium when the incumbent makes the first announcement.

Note that, in principle, the challenger's approval or disapproval of a particular sequence of announcements can in itself be informative, which implies that the characterization in Proposition 2 may not be unique. In particular, there may be other equilibria, in which for at least one sequence of announcements the probability distribution of $q$ is believed to differ from $p$ after the sequence has been agreed upon, which may lead to a completely different equilibrium characterization.

## 5. Conclusion

We consider a model of presidential debates with private information, which may be applicable in other contexts in which two individuals with conflicting interests decide whether to participate in some type of competition, taking into account that both the competition itself and their choice of whether to participate may reveal information about their qualities to an external decision maker. The results shed light on these situations, and in particular on the mutual effects between voters and candidates in presidential debates, by showing that the choice of whether to participate in a debate is uninformative since opting-out is not an option for the challenger.

**Appendix:**

Proof of Lemma 1i: Given that the debate is mandatory, by Bayes' rule:

$$(7)\ p(q')|w_m = \frac{p(q')\theta(q',q_I)}{\int_{q=0}^{\infty} p(q)\theta(q,q_I)\, dq}$$

and

$$(8)\ p(q')|l_m = \frac{p(q')(1-\theta(q',q_I))}{\int_{q=0}^{\infty} p(q)(1-\theta(q,q_I))dq} \quad \forall q' \in [0,\infty).$$

Therefore:

$$(9)\ \bar{q}|w_m = \int_{q=0}^{\infty} (p(q)|w_m) q\, dq = \frac{\int_{q=0}^{\infty} p(q)\theta(q,q_I)q\, dq}{\int_{q=0}^{\infty} p(q)\theta(q,q_I)\, dq}$$

and

$$(10)\ \bar{q}|l_m = \int_{q'=0}^{\infty} (p(q)|l_m) q\, dq = \frac{\int_{q=0}^{\infty} p(q)q(1-\theta(q,q_I))q\, dq}{\int_{q=0}^{\infty} p(q)(1-\theta(q,q_I))dq}.$$

By (7) and (8),

$$(11)\ p(q')|w_m > p(q')$$

$$\leftrightarrow$$

$$p(q')|l_m < p(q')$$

$$\leftrightarrow$$

$\theta(q', q_I) > \int_{q=0}^{\infty} p(q)\theta(q, q_I)dq.$

Given that by (1), $\theta(q_C, q_I)$ is concave in $q_C$, $\theta(\overline{q}, q_I) > \int_{q=0}^{\infty} p(q)\theta(q, q_I)dq (> 0)$. Therefore, since $\theta(q_C, q_I)$ is monotonically increasing in $q_C$, and by (1), $\theta(0, q_I) = 0$, by the Intermediate Value Theorem there exist $\hat{q} \in (0, \overline{q})$, such that $q' \gtreqless \hat{q} \leftrightarrow \theta(q', q_I) \gtreqless \int_{q=0}^{\infty} p(q)\theta(q, q_I)dq$. Considering (11), it follows that $q' \gtreqless \hat{q} \leftrightarrow p(q')|w_m \gtreqless p(q') \gtreqless p(q')|l_m$, which implies that:[21]

(12) $\overline{q}|w_m > \overline{q} > \overline{q}|l_m$.

By (1), (5) and (12), $E[G|0]|D_m = G((\overline{q}|l) - q_I) < G(\overline{q} - q_I)$, and when $q_C \to \infty$, $E[G|q_C]|D_m \to G((\overline{q}|w) - q_I) > G(\overline{q} - q_I)$.

Proof of Lemma 1ii: Substituting (9) and (10) into (6) results in:

(13) $EG|D_m$

$= (\int_{q=0}^{\infty} p(q)\theta(q, q_I)dq) G(\frac{\int_{q=0}^{\infty} p(q)\theta(q, q_I)qdq}{\int_{q=0}^{\infty} p(q)\theta(q, q_I)dq} - q_I) +$

$(\int_{q=0}^{\infty} p(q)(1 - \theta(q, q_I))dq) G(\frac{\int_{q=0}^{\infty} p(q)(1-\theta(q,q_I))qdq}{\int_{q=0}^{\infty} p(q)(1-\theta(q,q_I))dq} - q_I),$

---

[21] Note that since $\int_{q=0}^{\infty} p(q)dq = \int_{q=0}^{\infty} (p(q)|w_m) dq = \int_{q=0}^{\infty} (p(q)|l_m)dq = 1$, for each $q$ smaller than $\hat{q}$, the decrease in $p(q)|w_m$ relative to $p(q)$ must be fully compensated for by an increase in $p(q)|w_m$ relative to $p(q)$ for at least one value of $q$ greater than $\hat{q}$. A similar argument applies to $p(q)|l_m$. Therefore, $\int_{q=0}^{\infty} (p(q)|w_m) qdq > \int_{q=0}^{\infty} p(q) qdq > \int_{q=0}^{\infty} (p(q)|l_m) qdq$.

where by definition,

$$(14)\ G\left(\left(\int_{q=0}^{\infty} p(q)\theta(q,q_I)dq\right)\left(\frac{\int_{q=0}^{\infty} p(q)\theta(q,q_I)qdq}{\int_{q=0}^{\infty} p(q)\theta(q,q_I)dq} - q_I\right)\right.$$

$$\left. + \left(\int_{q=0}^{\infty} p(q)(1-\theta(q,q_I))dq\right)\left(\frac{\int_{q=0}^{\infty} p(q)\left(1-\theta(q,q_I)\right)qdq}{\int_{q=0}^{\infty} p(q)\left(1-\theta(q,q_I)\right)dq} - q_I\right)\right)$$

$$= G(\bar{q} - q_I).$$

Therefore, $EG|D_m > (<) G(\bar{q} - q_I)$ when G is convex (concave) over the interval $[(\bar{q}|l_m) - q_I, (\bar{q}|w_m) - q_I]$. QED

Proof of Lemma 2: By definition, $G\big((\bar{q}|C_{NP}) - q_I\big)(\in (0,1))$ is independent of $q_C$, where by (1) and (2), $E[G|q_C]|D$ is monotonically increasing in $q_C$.[22] Specifically, since $q_C \in [0,\infty)$, $E[G|q_C]|D \in \big[G((\bar{q}|l) - q_I), G((\bar{q}|w) - q_I)\big) (\subset [0,1])$. Therefore, if there exists a threshold level $q^* \in [0,\infty)$ that solves $G\big((\bar{q}|C_{NP}) - q_I\big) = E[G|q^*]|D$, then $q_C \gtreqless q^* \leftrightarrow E[G|q_C]|D \gtreqless G\big((\bar{q}|C_{NP}) - q_I\big)$.

However, it can be shown by contradiction that $q^*$ does not exist: Assume for now that there exists $q^* \in [0,\infty)$. Since it is commonly known that the challenger's response to P will be P iff $q_C \in [q^*,\infty)$,[23] $\overline{q}|w > q^*$, $\overline{q}|l > q^*$ and $\overline{q}|C_{NP} \leq q^*$, and therefore, by (2), $E[G|q_C]|D > G(q^* - q_I)$ for all $q_C$, and $G\big((\overline{q}|C_{NP}) - q_I\big) \leq G(q^* - q_I)$, which implies that $E[G|q_C]|D > G\big((\overline{q}|C_{NP}) - q_I\big)$ for all $q_C \in [0,\infty)$ including $q^*$, a contradiction.

---

[22] To see this, assume that the incumbent announces P. If the challenger announces NP, then $\bar{q}|c_{NP}$ and therefore also $G\big((\bar{q}|c_{NP}) - q_I\big)$ are both uniquely determined regardless of $q_C$, while if the challenger announces P, then $\bar{q}|w$ and $\bar{q}|l$ are both uniquely determined and therefore given (1) and (2), $E[G|q_C]|D$ is increasing in $q_C$.

Given that $E[G|q_C]|D$ is continuous in $q_C$ (and that $(\overline{q}|c_{NP})$ is independent in $q_C$) by the Intermediate Value Theorem, in equilibrium either $E[G|q_C]|D > G\big((\overline{q}|c_{NP}) - q_I\big)$ for all $q_C$ and therefore the challenger always responds to P with P or $E[G|q_C]|D < G\big((\overline{q}|c_{NP}) - q_I\big)$ for all $q_C$ and therefore the challenger always responds to P with NP.

Assume that the challenger always responds to P with NP. Given that $E[G|q_C]|D$ is increasing in $q_C$ and $(\overline{q}|c_{NP})$ is independent of $q_C$, if a challenger deviates and responds to P with P, then the quality of this challenger is believed to be ae least $q \geq 0$. Moreover, by (1) and (2), when $q_C \to \infty$, $E[G|q_C]|D \to G((\overline{q}|w) - q_I)$, where $\overline{q}|w > \overline{q} = \overline{q}|c_{NP}$ for any $\underline{q} \geq 0$[24] and therefore, given that $G$ is monotonically increasing in $\overline{q}^{el}$, $G((\overline{q}|w) - q_I) > G((\overline{q}|C_{NP}) - q_I)$, which implies that there exists a sufficiently large $q''$ above which $E[G|q_C]|D > G((\overline{q}|C_{NP}) - q_I)$ for all $q_C \geq q''$ and therefore a challenger with quality greater $q''$ deviates and responds to P with P, a contradiction.

---

[23] More precisely, the challenger is indifferent between P and NP when $q_C = q^*$.

Now assume that the challenger always responds to P with P. Given that $E[G|q_C]|D$ is increasing in $q_C$ and $(\overline{q}|c_{NP})$ is independent of $q_C$, if a challenger deviates by responding to P with NP, then the quality of this

---

[24] If $\underline{q}=0$, then $\overline{q}|w = \overline{q}|w_m$ where, by (12), $\overline{q}|w_m > \overline{q}$. If $\underline{q} > 0$, then the probability that the challenger's quality is $q'$ after the debate had been agreed upon and before it is held is $\frac{p(q')}{\int_{q=\underline{q}}^{\infty} p(q)dq}$ for all $q' \in [\underline{q}, \infty)$ and zero for all $q' \in [0, \underline{q})$, which implies that $p(q')|w = \frac{\frac{p(q')\theta(q',q_I)}{\int_{q=\underline{q}}^{\infty} p(q)dq}}{\frac{\int_{q=\underline{q}}^{\infty} p(q)\theta(q,q_I)dq}{\int_{q=\underline{q}}^{\infty} p(q)dq}} = \frac{p(q')\theta(q',q_I)}{\int_{q=\underline{q}}^{\infty} p(q)\theta(q,q_I)dq}$ for all $q' \in [\underline{q}, \infty)$ and $p(q')|w = 0$ for all $q' \in [0, \underline{q})$. Therefore, since, by definition, $\int_{q=\underline{q}}^{\infty} p(q)\theta(q,q_I)dq \leq \int_{q=0}^{\infty} p(q)\theta(q,q_I)dq$, then $\frac{p(q')\theta(q',q_I)}{\int_{q=\underline{q}}^{\infty} p(q)\theta(q,q_I)dq} \geq \frac{p(q')\theta(q',q_I)}{\int_{q=0}^{\infty} p(q)\theta(q,q_I)dq}$ which by (8) implies that $p(q)|w \geq p(q)|w_m$ for all $q \in [\underline{q}, \infty)$. Furthermore, the probability distribution of $q$ remains $p$ and therefore $\overline{q}|c_{NP} = \overline{q}$ when the challenger does not deviate.

challenger is believed to be at most $\underline{q} \leq \infty$, which implies that $(\overline{q}|C_{NP}) \leq \overline{q}$;[25] if he chooses not to deviate, then the probability distribution of $q$ remains $p$ and therefore, as shown above, there exists a sufficiently large $q''$ such that for all challengers for which $q_C > q''$, responding with P is better than responding with NP, which by the first part of the proof implies that no challenger deviates.   QED

Proof of Proposition 1: By Lemma 2, given that the incumbent announces P, the challenger will announce P for all $q_C$. Therefore, when the challenger announces P, regardless of the incumbent's announcement, the probability distribution of $q$ remains $p$, which implies that $EG|D = EG|D_m$. It follows that, when the challenger announces P, the incumbent's best response is to announce P when $EG|D_m < G(\overline{q} - q_I)$ and NP when $EG|D_m > G(\overline{q} - q_I)$, in which case by Fact 1 the challenger is indifferent between P and NP. However, at the end of the proof of Lemma 2 (including footnote 24) it is shown that if a challenger responds to P with NP, then her quality is $\underline{q} < \infty$

---

[25] In particular, either all challengers deviate and then $(\overline{q}|C_{NP}) = \overline{q}$ or only some deviate and then $\underline{q} < \infty$. If $\underline{q} < \infty$, then the probability that the challenger's quality is $q'$ is $\frac{p(q')}{\int_{q=0}^{\underline{q}} p(q)dq}$ for all $q' \in [0, \underline{q}]$ and zero for all $q' \in (\underline{q}, \infty)$. Since, by definition, $\int_{q=0}^{\underline{q}} p(q)dq \leq \int_{q=0}^{\infty} p(q)dq$, $\frac{p(q')}{\int_{q=0}^{\underline{q}} p(q)dq} \geq \frac{p(q')}{\int_{q=0}^{\infty} p(q)dq} = p(q')$ for all $q' \in [0, \underline{q})$, which implies that $\overline{q}|c_{NP} < \overline{q}$.

which implies that $\overline{q}|C_{NP} < \overline{q}$ and therefore the incumbent's best reply to the challenger's announcement of NP is to announce P. QED

Proof of Lemma 3i: By (1) and (2), $E[G|0]|D_m = G((\overline{q}|l_m) - q_I)$ and by (8), $\overline{q}|l_m > 0$. Therefore, by Lemma 1i and given that $G$ is monotonically increasing in $\overline{q}^{el}$, $G(-q_I) < E[G|0]|D_m < G(\overline{q} - q_I)$.

When $q_C \to \infty$, by (1) and (2) $E[G|q_C]|D_m \to G((\overline{q}|w_m) - q_I)$, and, by definition, $G(q_C - q_I) \to 1$. Given Lemma 1i and (9), it follows that when $q_C \to \infty$, $G(q_C - q_I) > E[G|q_C]|D_m > G(\overline{q} - q_I)$.

Therefore, since $G$ is continuous and monotonically increasing in $\overline{q}^{el}$, by the Intermediate Value Theorem, there exist $q_L \in (0, \overline{q})$ and $q_H \in (\overline{q}, \infty)$ such that $|E[G|q_C]|D_m - G(q_C - q_I)| > |G(\overline{q} - q_I) - G(q_C - q_I)|$ for all $q_C < q_L$ and $q_C > q_H$.

Proof of Lemma 3ii: Note that $|G(\overline{q} - q_I) - G(q_C - q_I)| \approx 0$ when $q_C \approx \overline{q}$, Furthermore, it was shown above that $E[G|q_C]|D_m > G(q_C - q_I)$ when $q_C$ is sufficiently small and $E[G|q_C]|D_m < G(q_C - q_I)$ when $q_C$ is sufficiently large and therefore in both cases $|E[G|q_C]|D_m - G(q_C - q_I)| > 0$. QED

Proof of Proposition 2: Fix the order of announcements such that the incumbent moves first. Since her announcement is uninformative by definition, the challenger's best response remains the same as in the simultaneous case, implying that in equilibrium the probability distribution of $q$ remains $p$ and therefore for any pair of announcements $EG$ has the same value as in the simultaneous case, implying that Proposition 1 satisfies Perfect Bayesian Equilibrium of the subgame in which the incumbent is the leader.[26]

Alternatively, fix the order of announcements such that the challenger announces first. Then, the incumbent responds to P with P when $EG|D < G(\overline{q}|I_{NP} - q_I)$, which by Fact 1 implies that $EG|D < G(\overline{q} - q_I)$. It is therefore commonly known that a debate is held iff the challenger announces P when $EG|D < G(\overline{q} - q_I)$ and by Lemma 2 she then announces P for all $q_C$, which implies that a debate is held when $EG|D < G(\overline{q} - q_I)$, where $EG|D = EG|D_m$. As a result, it is commonly known that the incumbent prefers that a debate not be held when $EG|D_m > G(\overline{q} - q_I)$ and therefore in view of Fact 1 any announcement made by the challenger in this case is followed by an announcement of NP by the incumbent, which implies that any pair of announcements except {P,P} satisfies Perfect Bayesian Equilibrium of this subgame when $EG|D_m > G(\overline{q} - q_I)$.

Hence, Proposition 1 satisfies Perfect Bayesian Equilibrium of all subgames in which the sequence of announcements is fixed (simultaneous,

---

[26] Note that since by Fact 1, the challenger's response to NP can be either NP or P, both pairs of announcements (NP,NP) and (NP,P), satisfy a Perfect Bayesian Equilibrium when the incumbent makes the first announcement and $EG|D_m > G(\overline{q} - q_I)$ (this also implies that $EG$ is then different from the simultaneous case when the announcements are (NP,P)).

.

challenger announces first or incumbent announces first). Therefore, given that the probability distribution of *q* remains *p* regardless of the sequence of announcements after such a sequence has been agreed upon, the equilibrium expected probability of winning the elections for both candidates is independent of the sequence of announcements. It follows that the beliefs, namely that the probability distribution of *q* remains *p* after the sequence of announcements has been agreed upon for all sequences, are consistent with the candidates' strategies. Given these beliefs, the challenger has no reason to deviate by accepting a particular sequence of announcements only for a subset of *q*, since for a given $q_C$, $E[G|q_C]$ is the same for all sequences. QED